\documentclass[12pt]{iopart}
\usepackage{epsf}

\begin{document}

\title[]{A Non-extensive Equilibrium analysis of $\pi^{+}$ $p_{T}$ Spectra at RHIC}

\author{D.D. Chinellato$^{1}$, J. Takahashi$^{1}$, I. Bediaga$^{2}$}
\address{$^{1}$ Universidade Estadual de Campinas (UNICAMP), Campinas, Brazil\\$^{2}$ Centro Brasileiro de Pesquisas Fisicas, Rua Xavier Sigaud 150, 22290-180 – Rio de Janeiro, RJ, Brazil }
\ead{daviddc@ifi.unicamp.br}

\begin{abstract}
By analyzing the dynamical properties of particle production in
relativistic heavy ion collisions, it is possible to characterize the
final stage of the equilibration process occurring in the collision fireball. In this
work, we use the Hagedorn model coupled with non-extensive statistics to evaluate the transverse momentum spectra of positive pions
for various event centrality classes in Au+Au collisions at RHIC
with center-of-mass energies of 62.4GeV/A and 200GeV/A. We
find that, by assuming an energy distribution that incorporates
particle correlations, it is possible to explain the entire $\pi^{+}$ $p_{T}$
spectrum as measured by RHIC.
We find that spectra from central
collisions, when compared to peripheral collisions, are consistent
with a system that has smaller values of the non-extensivity parameter ``q''
 and higher values of temperature. Comparison between different beam energies also shows a
variation of the ``q'' parameter. The result is discussed using the
interpretation that the ``q'' parameter is a measure of particle
correlations within the system. Under this assumption, our results
show that more central collisions are consistent with a system with
less particle correlations.\end{abstract}

\section*{Introduction}
Relativistic Heavy Ion collisions have been studied for
their potential in exploring many aspects of the QCD phase diagram. The 
particle production, as measured by experiments such as the Relativistic Heavy Ion
Collider (RHIC) at the Brookhaven National Laboratory, exhibits transverse momentum
spectra that are divided into three regions described by different physical models \cite{exthechal}.
For instance, particles with relatively low momenta ($<1.5GeV/c$), which exhibit an exponential
distribution, can be understood to be thermalized according to Boltzmann 
statistics. The intermediate $p_{T}$ range, $1.5GeV<p_{T}<6GeV/c$, might be considered as coming
from quark coalescence and recombination, while the highest momenta $p_{T}>6GeV/c$ 
could originate from parton fragmentation. Thus, transverse momentum spectra are usually
explained differently according to the momentum range considered.

It would be of interest to find a single physical description for measured
momentum spectra over the entire $p_{T}$ range. This has been attempted before with different approaches. For
instance, a framework of percolation of strings can be used to find a universality in the transverse momentum 
spectra \cite{pajares}. Alternatively, temperature fluctuations have been shown to generate spectra compatible
to the ones measured experimentally \cite{wilk}. 
In this work, we chose to use a non-extensive statistical approach \cite{Tsallis1,Hagedorn1,biyajima}. In this 
formalism the entropy is no longer additive. When computed in the canonical ensemble, the occupation 
probabilities are given by

\begin{equation}
\label{one}
p_{i}=\frac{1}{Z_{q}}\left[1-\beta\left(q-1\right)\epsilon_{i}\right]^{\frac{1}{q-1}}
\end{equation}
with $Z_{q}$ being
\begin{equation}
\label{two}
Z_{q}=\sum_{i-1}^{W}\left[1-\beta\left(q-1\right)\epsilon_{i}\right]^{\frac{1}{q-1}}
\end{equation}where $p_{i}$ is the occupation probability of a state with energy $\epsilon_{i}$, $W$ is 
the number of accessible states, $\beta$ is a real number and ``$q$'' is the non-extensivity parameter, so called because the
 composition rule for entropy deviates from additiveness by a mixed entropy term proportional 
to $(1-q)$. These expressions reduce to the Boltzmann  occupation probabilities in the limit
 $q\rightarrow 1$, where entropy also becomes additive as in classical statistical theory. 

The probabilities given by equation \ref{one} therefore give us an energy distribution, but it is 
still necessary to relate energy to transverse momentum. To do this, we assume a thermal equilibrium 
of a relativistic gas, as in the Hagedorn model, where energy is distributed according to equation \ref{one}.
One can then determine the momentum distribution to be \cite{Bediaga}

\begin{equation}
\label{three}
\frac{1}{\sigma}\frac{d\sigma}{dp_{T}}=cp_{T}\int_{0}^{\infty}dp_{L}\left[1-\frac{1-q}{T}\sqrt{p_{L}^{2}+\mu^{2}}\right]^{\frac{q}{1-q}}
\end{equation}

Here, $T$ denotes a real number equal to $1/\beta$ which is related to the slope of the resulting curve, 
$p_{L}$ is the longitudinal momentum and $p_{T}$ is the transverse momentum of the produced particles, and 
$\mu^{2}=m_{0}^{2}+p_{T}^{2}$, where $m_{0}$ is the rest mass, is the so called transverse mass of the
produced particles.

This function can be used numerically to adjust transverse momenta. It is, however,
of crucial importance to understand the fit parameters in order to gain physical knowledge of
the dynamics at hand. 

\section*{Dynamical Model}

A dynamical interpretation for the $q$-parameter has been suggested in reference \cite{KodamaKoide}
and proposes a discrete dynamical model which can be simulated easily. Consider a system consisting of 
$N$ particles, each having energy $E_{i}$ in multiples of some energy quantum, $\Delta E$. The system evolves 
in discrete time steps in which two particles $i$ and $j$ are randomly chosen to be updated and one of the following 
two updates takes place:
\begin{itemize}
\item With probability $1-r$, the two particles exchange one energy quantum, i.e. $E_{i}=E_{i}\pm\Delta E$ and $E_{j}=E_{j}\mp\Delta E$. 
If the choice is such that a negative energy should arise, nothing happens and this step is skipped.
\item With probability $r$, the two particles are set to the energy of the lowest original energy, i.e.
$E_{i}=E_{j}=Min(E_{i},E_{j})$, and the excess energy is passed on to a third randomly chosen particle, effectively
introducing a two-particle correlation. 
\end{itemize}

It is therefore possible, by changing the $r$ parameter in the model, to increase or decrease the degree
of correlation in the dynamics. We simulated such systems using $N=100$ in a system saturated
with energy, in which the average energy per particle is $\left<E\right>=180$. In this case, at least
$10^{6}$ time steps are required in order to attain an energy distribution curve that is stationary. It is 
noteworthy that different $N$ values change this relaxation time, as more iterations will be needed to attain
equilibrium if more particles are present.

\begin{figure}
\centerline{\epsfxsize 6.2in \epsffile{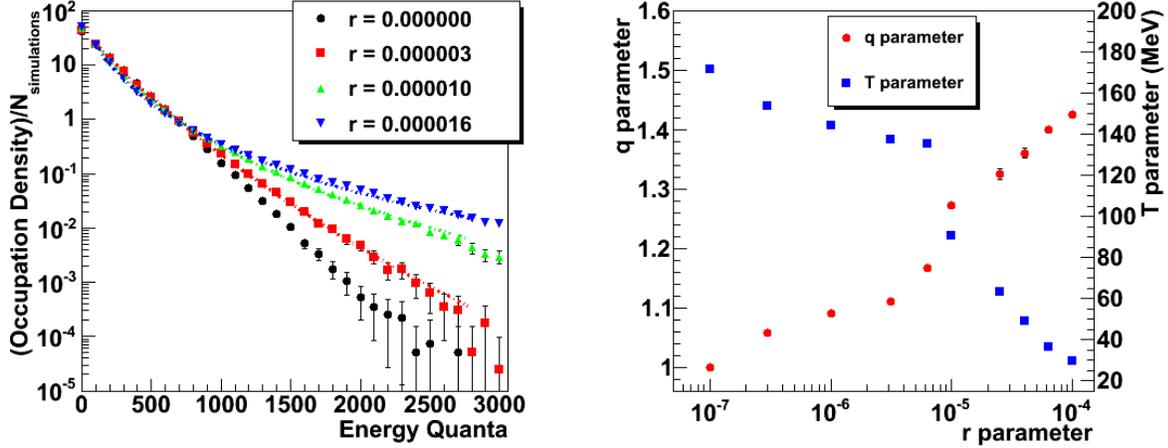}}
\caption{\label{label1}  a) Tsallis distribution fit to spectra generated using different {\it r} values
in discrete dynamical model.
b) Obtained $q$ and $T$ fit parameters according to the {\it r} value in the dynamical model. }
\end{figure}

This model produces energy distributions that are similar to Tsallis' energy distribution curves and can 
be fitted with equation \ref{one}, shown in Fig. \ref{label1}a. The fit parameters indicate that the $r$ 
value and the adjusted $q$ value are closely related (Fig. \ref{label1}b) and the $q$ value is indicative
of a higher degree of correlation in the dynamics. Fig. \ref{label1}b also
shows that the $T$ and $q$ parameters are anti-correlated, which can be seen as a reflection of conservation of energy. Populating
higher energies with a small amount of particles implies taking this energy from the bulk of the other particles, effectively
increasing the slope parameter and decreasing the $T$ parameter. Other configurations varying $N$ and $\left<E\right>$ were 
also studied and in all cases we observed the same qualitative behaviour for different $r$ values. 

\begin{figure}
\centerline{\epsfxsize 6.2in \epsffile{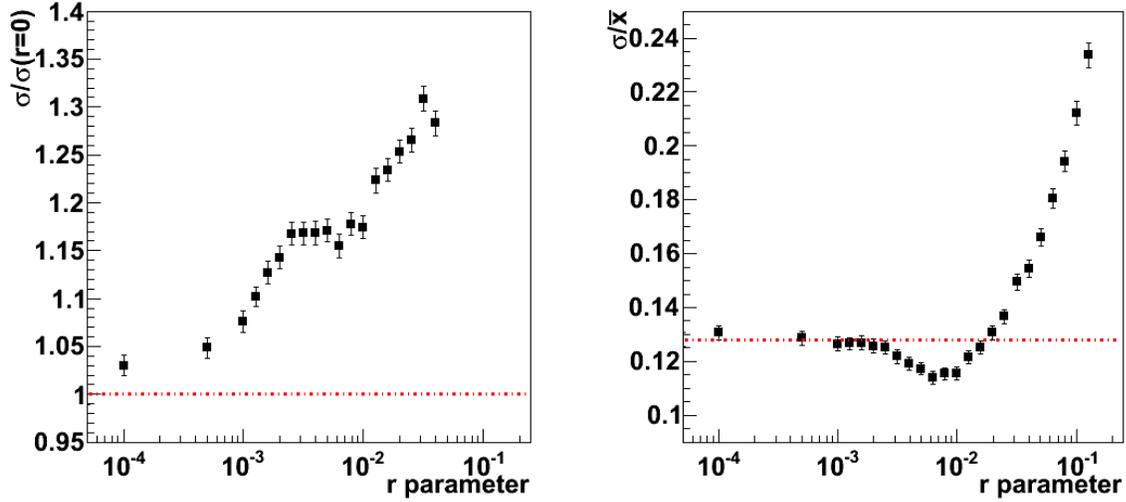}}
\caption{\label{label1a}  Systematic study of event-by-event fluctuations, as measured by width of gaussian 
fits to energy occupancy using discrete dynamical model. a) shows data normalized by reference width and b) shows
data normalized according to gaussian mean. The dashed line corresponds to the reference value for $r=0$.}
\end{figure}

This toy model also makes the study of event-by-event energy fluctuations possible. By simulating many 
independent systems and looking at the energy occupation of the $2-15$ energy quanta bin and fitting these data
with a gaussian curve, it can be seen that the fluctuations increase by roughly $20-25\%$ (Fig. \ref{label1a}a)
when increasing the correlation parameter $r$ in the model. If we normalize the gaussian widths by the mean value,
it can be seen (Fig. \ref{label1a}b) that this increase in the $\sigma$ values is not due to the central 
limit theorem for $r>10^{-2}$, as this would require that the increase be proportional to the mean value.

\section*{Comparing with Experimental Data}

\begin{figure}
\centerline{\epsfxsize 6.2in \epsffile{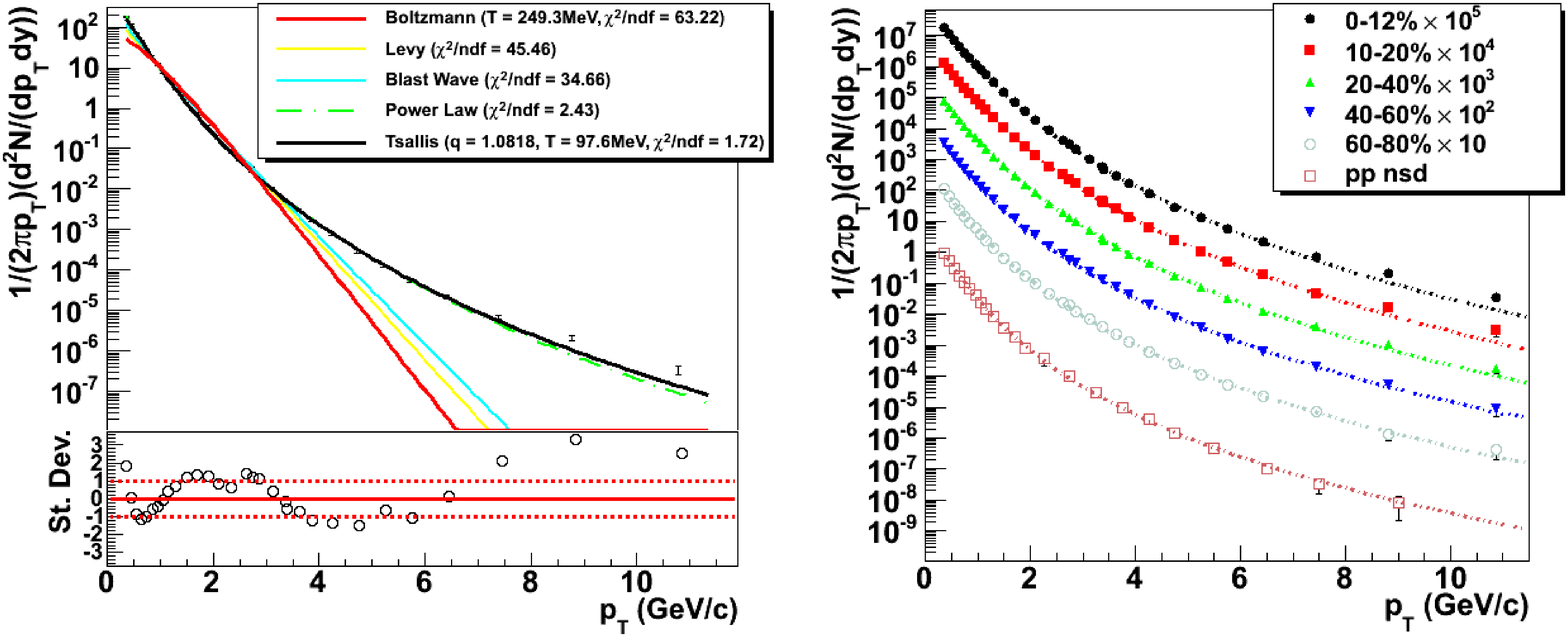}}
\caption{\label{label2}  a) Comparison of full-range fits to RHIC $\pi^{+}$ data at 200GeV.
b) Hagedorn model with Tsallis energy distribution fits accross several centrality bins and
p+p data from RHIC at 200GeV.}
\end{figure}

We have looked at the $\pi^{+}$ momentum distribution as measured and published by the STAR experiment from pions
from Au+Au and p+p collisions (see \cite{200GeVData} and
\cite{62GeVData}) and fitted the Hagedorn gas model coupled with Tsallis' distribution, as in eq. \ref{three}. This can be 
readily compared to other full-range candidate functions, as seen in Fig. \ref{label2}a. The best $\chi^{2}$ value obtained
is from the Hagedorn+Tsallis model. 

This approach can be used systematically accross different centrality bins, as well as with p+p data, as shown in Fig. \ref{label2}b. 
It is remarkable that the $\chi^{2}$ values are good throughout all these datasets. Using data measured with a beam energy of 62.4GeV also
results in the Hagedorn+Tsallis function having the best $\chi^{2}$ values throughout all centrality bins studied.

\begin{figure}
\centerline{\epsfxsize 6.1in \epsffile{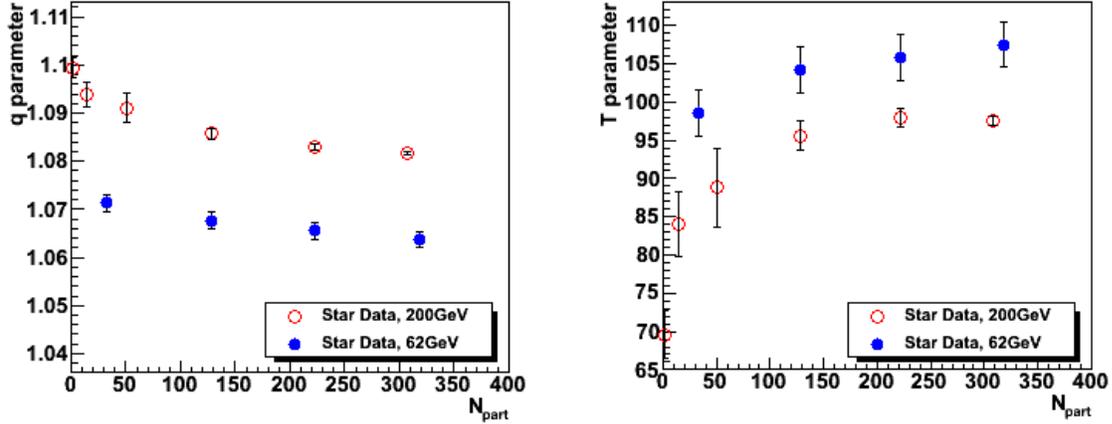}}
\caption{\label{label3}  $q$ and $T$ fit parameters obtained for 200GeV Au+Au, p+p and 62.4GeV Au+Au data, according to number of participants in 
the collision, calculated with optical glauber model. Larger number of participants correspond to more central collisions.}
\end{figure}

The dependence of the $q$ and $T$ parameters from these fits show that the more central collisions always exhibit a lower $q$-value
when fitting, as well as a higher $T$ parameter (Fig. \ref{label3}). The p+p data fits well in the systematic trend. Considering
the Dynamical Model introduced in the previous section, where it was proposed that $q$ could be used as a measure of the correlation
in the dynamics, the results shown in Fig. \ref{label3} suggest that more central collisions have dynamics in which particles are less correlated. If the main source of correlation
is assumed to be jet production, this could be an indication of jet quenching. The slight decrease in the $T$ parameter is also consistent
with an increased $r$ value in the dynamical model. 

Comparing the fit parameters for the two different datasets measured with two beam energies, it can be seen that the higher beam energies exhibit 
a higher $q$-value as well as lower $T$ parameters. Higher $q$ values, using the same dynamical model analogy, translate to higher $r$ value and
a higher jet formation. 

\begin{figure}
\centerline{\epsfxsize 6.1in \epsffile{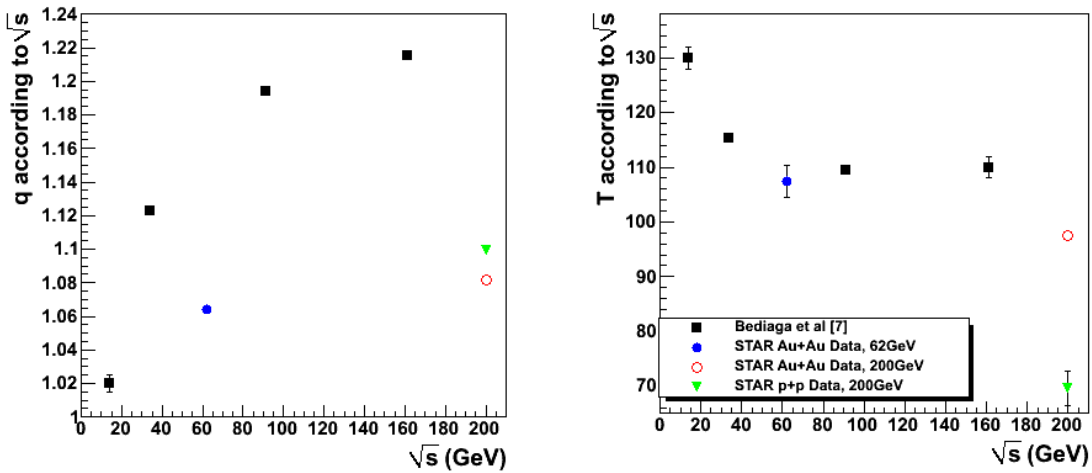}}
\caption{\label{label4}  Comparison to fit results published for $e^{+}e^{-}\rightarrow hadrons$ by Bediaga et Al \cite{Bediaga}. Only most central data from heavy ion collisions have been plotted.}
\end{figure}

It is noteworthy that this analysis has been done before (\cite{Bediaga}) with $e^{+}e^{-}\rightarrow hadrons$. Our results can be compared to the 
previous as in Fig. \ref{label4}, and even though the two physical processes are very different, in both cases the Hagedorn Gas Model coupled with
Tsallis statistics is quite effective in describing the spectra obtained. The general trend of higher $q$-values and lower $T$-values is the same in
both analyses. 

\section*{Conclusion}

We have shown that, by using the Hagedorn gas model with energies distributed according to Tsallis statistics, the full-range transverse momentum
spectra from heavy ion and p+p collisions can be described. It is remarkable that this approach works well for both heavy ions and p+p, 
and throughout the entire measured range of transverse momenta.

A preliminary interpretation of the fit parameters indicates that there are less correlations in the dynamics of more central collisions, as well
as a higher slope parameter $T$. If we take our dynamical model and map its parameters to $q$ and $T$, we see that this is consistent with a smaller
$r$ value. If we interpret correlations in the dynamics as coming mainly from jets, this
indicates that in more central collisions there are proportionally less jets, i.e. jet quenching. A higher energy density has also been measured before 
for more central AuAu collisions at RHIC \cite{Jun}. In addition, when comparing transverse momentum data from collisions at different beam energies, we see higher 
$q$ parameters, which, when mapped to higher $r$ values and using the same analogy, is an indication of dynamics in which more jets are produced. 

Further studies are warranted to fully understand the fit parameters as well as the physics that lies in the full-range transverse momentum spectra 
considered as such. It would also be of interest to investigate the spectra of other particle species using this same framework. Other frameworks have already been used to do this \cite{zhangbu}, where the blast wave model is used to describe various particle spectra. 

Finally, it is noteworthy that the transverse momentum range considered in this work is relatively large compared to the available transverse momentum data for other particle species due to experimental limitations. Our investigations show that having data all the way up to roughly $4-5GeV/c$ is best for the parameters to converge ideally.  

\section*{Acknowledments} 
We wish to thank Funda\c{c}\~ao de Amparo a Pesquisa do Estado de S\~ao Paulo, FAPESP,
Brazil for the support. We also thank Prof. Dr. Grzegorz Wilk for discussions and input.

\end{document}